\documentclass[runningheads]{llncs}
\usepackage[T1]{fontenc}
\usepackage{graphicx}
\usepackage{amsmath}
\usepackage{algorithm}
\usepackage[noend]{algpseudocode}
\usepackage{url}
\usepackage{hyperref}
\usepackage{color}

\begin{document}
\title{Demonstrating Narrative Pattern Discovery from Biomedical Literature}
\author{Hermann Kroll\inst{1}\orcidID{0000-0001-9887-9276} \and Pascal Sackhoff\inst{1}\orcidID{0009-0005-3095-9794} \and Bill Matthias Thang\inst{1}\orcidID{0009-0006-8321-8479} \and
Christin Katharina Kreutz\inst{2,3}\orcidID{0000-0002-5075-7699} \and
Wolf-Tilo Balke\inst{1}\orcidID{0000-0002-5443-1215}}
\authorrunning{Kroll et al.}
\institute{
TU Braunschweig, Braunschweig, Germany\\
\email{krollh@acm.org, balke@ifis.cs.tu-bs.de} 
\and
TH Mittelhessen - University of Applied Sciences, Gießen, Germany \and
Herder Institute, Marburg, Germany\\
\email{ckreutz@acm.org}\\
}
\maketitle              % typeset the header of the contribution
\begin{abstract}
Digital libraries maintain extensive collections of knowledge and need to provide effective access paths for their users. For instance, PubPharm, the specialized information service for Pharmacy in Germany, provides and develops access paths to their underlying biomedical document collection. In brief, PubPharm supports traditional keyword-based search, search for chemical structures, as well as novel graph-based discovery workflows, e.g., listing or searching for interactions between different pharmaceutical entities. This paper introduces a new search functionality, called narrative pattern mining, allowing users to explore context-relevant entities and entity interactions. We performed interviews with five domain experts to verify the usefulness of our prototype.

\end{abstract}

\section{Introduction}
Digital libraries maintain extensive collections of knowledge and need to provide effective access paths for their users. Ideally, two types of searches should be supported: precise search for relevant material and exploratory search~\cite{DBLP:journals/jodl/KrollPKKRB24}. In this paper, we focus on exploratory search to allow users to find new and interesting ideas for their own work.
For instance, the connected papers service\footnote{\url{https://connectedpapers.com}} allows users to explore the connection between different research articles, e.g., who cites whom or what is adjacent to a certain paper. This way users may find new and interesting articles for their own work. 

We, as the specialized service for Pharmacy in Germany, build upon that idea. The biomedical/pharmaceutical domain is an entity-centric one, e.g., research focuses around certain drugs, diseases, targets, methods and more; see for instance a PubMed query log analysis~\cite{herskovic2007pubmedqueryanalysis} or entity-centric services like PubTator~\cite{wei2019pubtatorcentral}. That is why we developed a new entity-centric search functionality for our platform, which we are describing in the following. The main idea is that users start their search with a set of relevant entities for their own work. Then, the service first retrieves documents that include these entities and second, mines patterns between the given and other context-relevant entities. All information is then visualized as a network so that users can explore context-relevant entities and entity interactions, so called \textit{narrative patterns}. A click on a network's edge forwards users to corresponding literature supporting the selected entity-entity interaction. While our Narrative Discovery System has been published~\cite{DBLP:journals/jodl/KrollPKKRB24}, this paper introduces our \textit{narrative pattern}-driven discovery method and prototye.

\begin{figure}[t]
    \centering
    \includegraphics[trim={0 0.5cm 0 0}, clip, width=1\linewidth]{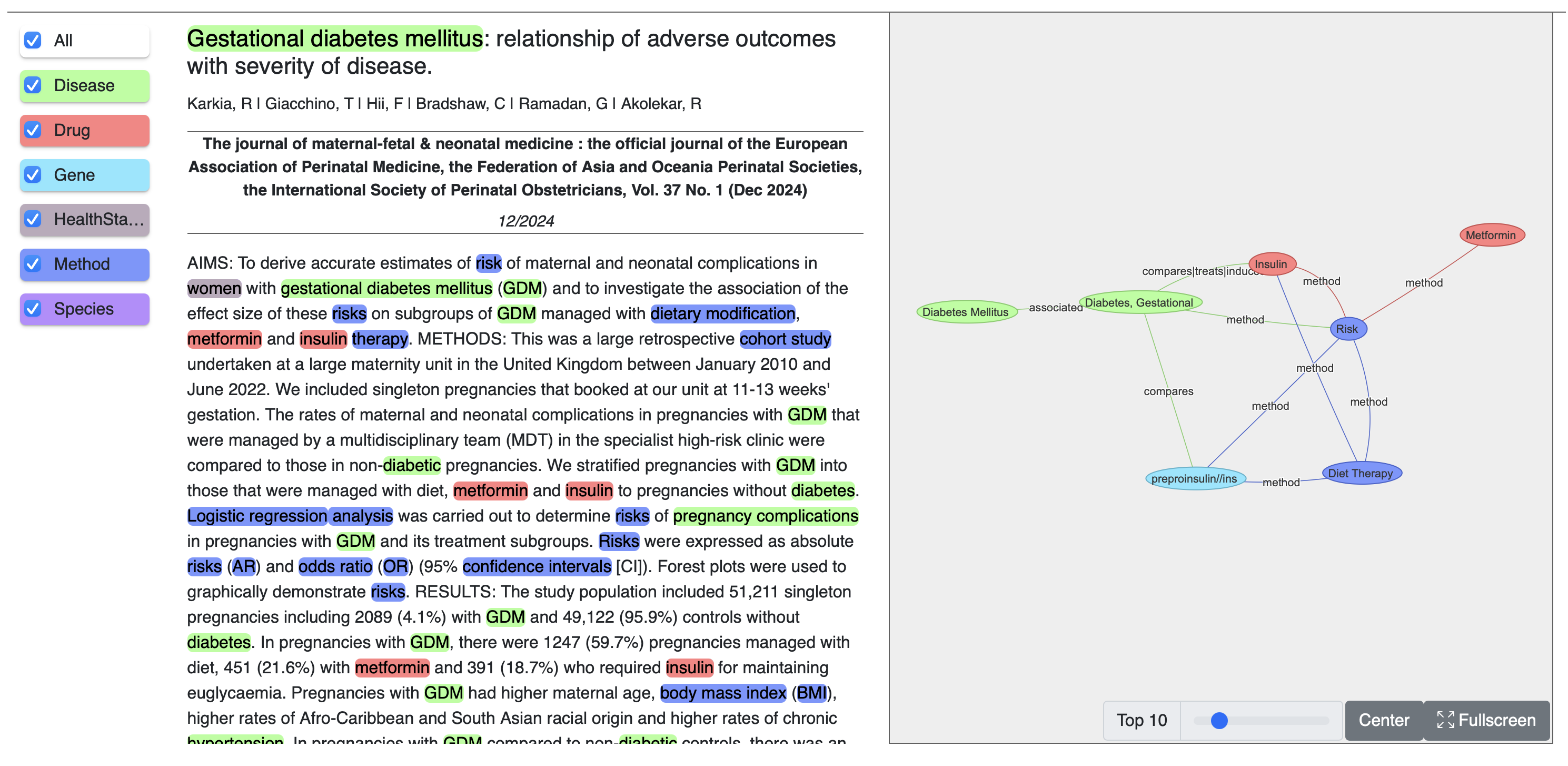}
    \caption{Document Visualization: The left side shows the document text. Detected entities are highlighted in corresponding colors. A UI selection box on the left side allows to show or hide certain entity types. The right side depicts extracted interactions between entities as a labeled and colored graph.}
    \label{fig:document}
\end{figure}

\section{Narrative Discovery System}
\label{chap:narrativeservice}
PubPharm\footnote{\url{https://www.pubpharm.de}}, the specialized information service for Pharmacy in Germany, aims to provide effective and innovative access paths to the pharmaceutical literature for our research community. 
In the past, we proposed and implemented a discovery system for narrative information access~\cite{DBLP:journals/jodl/KrollPKKRB24}. 
The system called the Narrative Service\footnote{\url{https://narrative.pubpharm.de}} allows users to formulate their information needs as graph patterns, i.e., interaction patterns between entities. 
This way, users may search for literature stating that \textit{Metformin is used to treat diabetes mellitus in adult patients}. 
In addition, variables can be used to explore the literature, e.g., \textit{any drugs} used to treat diabetes mellitus in adult patients. 
The service is capable of answering these queries through graph pattern matching. 
The Drug Overviews\footnote{\url{https://narrative.pubpharm.de/drug_overview}} service extends our discovery system by allowing users to generate overviews about specific drugs, i.e., known interactions like therapies, target interactions, administrations, and more. 
When users click on some information, they are directed to a corresponding search in the Narrative Service, e.g., searching for literature about a certain drug-disease therapy. 

To enable graph pattern matching, the relevant biomedical literature is transformed into document graphs, i.e., relevant biomedical entities are detected, and their interactions are extracted from texts. 
Instead of building a single knowledge graph, we decided to transform each text into a small document graph and keep these graphs separated.
Reasons for this decision were that 1) user queries are still answered by retrieving relevant literature instead of short answers and 2) the validity of statements is ensured~\cite{DBLP:conf/ercimdl/KrollKNMB20}, i.e., the system retrieves the corresponding context in which a statement is valid.
A visualization of an enriched document in our system is shown in \autoref{fig:document} or online\footnote{\url{https://narrative.pubpharm.de/document/?document_id=38844413&data_source=PubMed}}.
Entities were detected by performing a dictionary-based entity linking against existing vocabularies and by using existing biomedical annotation tools like GNormPlus~\cite{wei_gnormplus_2015}, TaggerOne~\cite{leaman_taggerone_2016}, and PubTator Central~\cite{wei2019pubtatorcentral}. 
Statements were extracted by deploying PathIE, a method for extracting statements via the grammatical structure of sentences, and by extracting association statements if two entities co-occur within the same sentences.
These methods are part of our extraction toolbox~\cite{DBLP:conf/jcdl/KrollPB21}, which we described~\cite{DBLP:phd/basesearch/Kroll23} and analyzed~\cite{DBLP:journals/jodl/KrollPPB24} comprehensively.
In brief, most of our self-developed methods do not rely on supervision and thus bypass the need for training data, but they only come with a moderate extraction quality.
The code for our discovery platform and the extension for this paper is available at GitHub\footnote{\url{https://github.com/HermannKroll/NarrativeIntelligence/}} and SoftwareHeritage\footnote{SoftwareHeritage ID: \href{https://archive.softwareheritage.org/swh:1:dir:5b87566505d9f3ad0837cc91f105ee163515ec3d}{swh:1:dir:5b87566505d9f3ad0837cc91f105ee163515ec3d}}.
Our system (as of May 2025) includes 38M PubMed/Medline articles.

The Narrative Service itself provides precise literature searches due to graph-based queries and, thus, entity-interaction-aware searches~\cite{DBLP:journals/jodl/KrollPKKRB24}. 
A keyword-based search functionality assists users in formulating graph queries~\cite{DBLP:conf/jcdl/KrollKSB23}.
Exploratory searches are, as of now, supported by using variables in queries or using the Drug Overview functionality.
This paper contributes a new access path for our service that allows users to discover entity interactions in contexts.

\section{Mining Narrative Patterns}
Entity interactions play a central role within the biomedical literature.
The goal of our narrative pattern mining here is to allow an exploration of the literature, i.e., visualizing and thus summarizing what is known and often described between a set of searched entities.
With that, the system can shed light on context-relevant entities and interactions between them, so that users can explore the entities' neighborhood. 
This kind of exploration should assists users in understanding the relationships between biomedical entities and possibly discover new relevant entities to the users' information needs.

\paragraph{System Architecture.} 
In brief, our service expects a list of searched entities as its input. 
The output is a graph pattern that 1) puts the searched entities in relation and 2) adds more entities that play a central role in the searched entities' contexts.
Therefore, we first identify relevant documents, retrieve the document graphs, score the graphs' edges, and sort these edges by their final score.
Users can then select how many edges shall be displayed.

\paragraph{Identifying Relevant Documents.}
Our goal is to support users in exploring the literature by showing what is written about the set of searched entities. 
We, therefore, decided to extract these patterns from documents that include all of the searched entities. 
This way, only information appearing in a context (a document) that includes all searched entities is considered. 

Users enter a list of strings.
Each string represents a search for entities in our system. 
Each of these strings has to be translated into entities from our vocabulary. 
Therefore, we use the following translation paradigm: 
Suppose a user types the string \textit{diabetes melli} in the search but does not complete their string insertion yet.
All entities that include both the terms \textit{diabetes} and \textit{melli} in one of their synonyms are valid translations, e.g., \textit{diabetes mellitus}, \textit{diabetes mellitus type 1}, \textit{diabetes mellitus type 2}, and many more.
These valid translations are then suggested to users during the input as keywords.
A user can pick one of the suggestions, e.g., \textit{diabetes mellitus} as keyword\footnote{Note that selecting \textit{diabetes mellitus} as keyword will still also translate the keyword to \textit{diabetes mellitus type 1} etc.} or finish their typing and lock in their inserted string as a keyword.
With that, we can translate the user's input keyword into a set of entities.
Next, we use an inverted index to retrieve document IDs in which a particular entity has been detected.
This gives us the set of documents relevant for a specific keyword.

If a user enters multiple keywords, the translation is conducted for all keywords. 
Then we compute the intersection between those sets of documents relevant for \textit{single} keywords to identify documents fitting \textit{all} entered keywords.

\paragraph{Scoring Edges.}
Next, we retrieve the document graphs for the retrieved document IDs fitting all keywords and mine the actual narrative pattern. 
Combining all graphs and showing the resulting one to the user would likely to be overstraining them.
For instance, when searching for entities like \textit{diabetes mellitus} and \textit{metformin}, thousands of documents are retrieved. 
The resulting graph would then also include hundreds of different statements.
We tackled this problem in two ways:
1) We score each graph edge so that only the most important ones are shown to the users. Users can control how many edges should be visualized at once.
2) We only show edges that are incoming or outgoing from one of the user's searched entities so that these entities are put into the focus of the generated narrative pattern.
We score graph edges as follows, which proved to be effective for ranking and recommending documents in our discovery system~\cite{DBLP:journals/corr/abs-2412-15229,DBLP:journals/corr/abs-2412-15232}:
$    \textit{score}(e, d) = \textit{tf-idf}(e, d) * \textit{coverage}(e, d) * \textit{confidence}(e, d)
$

The scoring function takes a graph edge $e$ and its corresponding document $d$ as its input and returns a numeric score.
In brief, tf-idf stands for term-frequency inverse-document-frequency and prefers edges that appear often within $d$ but rarely within the whole collection.
Coverage favors edges that include entities used across the document (and not just at the beginning or end of some text). 
Confidence boosts edges with high extraction confidence, i.e., the extraction method extracted the edge with high confidence.
For more details we refer the reader to our prior works~\cite{DBLP:journals/corr/abs-2412-15229,DBLP:journals/corr/abs-2412-15232} or our actual implementation.

The score function allows us to score each document graph edge.
For the narrative pattern, we compute the union of all retrieved graphs. 
Then, we score each of its edges as follows.
Let $D$ be the set of retrieved documents relevant for all keywords and $e$ be an edge of the narrative pattern, we sum the scores for this edge in every document from $D$ for our final score: $\textit{fscore}(e) = \sum_{d \in D} \textit{score}(e, d)$.
\textit{score} returns 0 if the edge $e$ is not present in document $d$.

Our final Algorithm is shown in~\ref{alg:PatternScoring}. 
First, it retrieves relevant documents and then sums up the scores for each edge in every retrieved document. 
The edges are then sorted in descending order with regard to their scores.

\begin{algorithm}[t]
\caption{Mining Narrative Patterns from Document Graphs}
\label{alg:PatternScoring}
\begin{algorithmic}[1]
\State \textbf{Input}: list(set(entities translated from each keyword)) \textbf{Output}: a ranked list of documents 
\State $relDocPerKeyword$ = list()
\For{$entitiesForKeyword \in$ listOfEntitiesPerKeyword}
    \State $relDocPerKeyword$.append(getRelevantDocs($entitiesForKeyword$))
\EndFor
\State $intersectRelDocs$ = set.intersection(*relDocPerKeyword)
\State edge2score = dict()
\State docs = retrieveDocumentData($intersectRelDocs$)
\For{$d \in$ docs}
\For{$e \in$ d.edges}
\If{$e.subject \in \text{entities} \lor e.object \in \text{entities}$}
    \State edge2score[s] += score(e, d)
\EndIf
\EndFor
\EndFor
\State edge2score = sortByScoreDescending(edge2score)

\State \textbf{return} edge2score

\end{algorithmic}
\end{algorithm}

\begin{figure}[t]
    \centering
    \includegraphics[trim={0 0.5cm 0 0}, clip, width=0.9\linewidth]{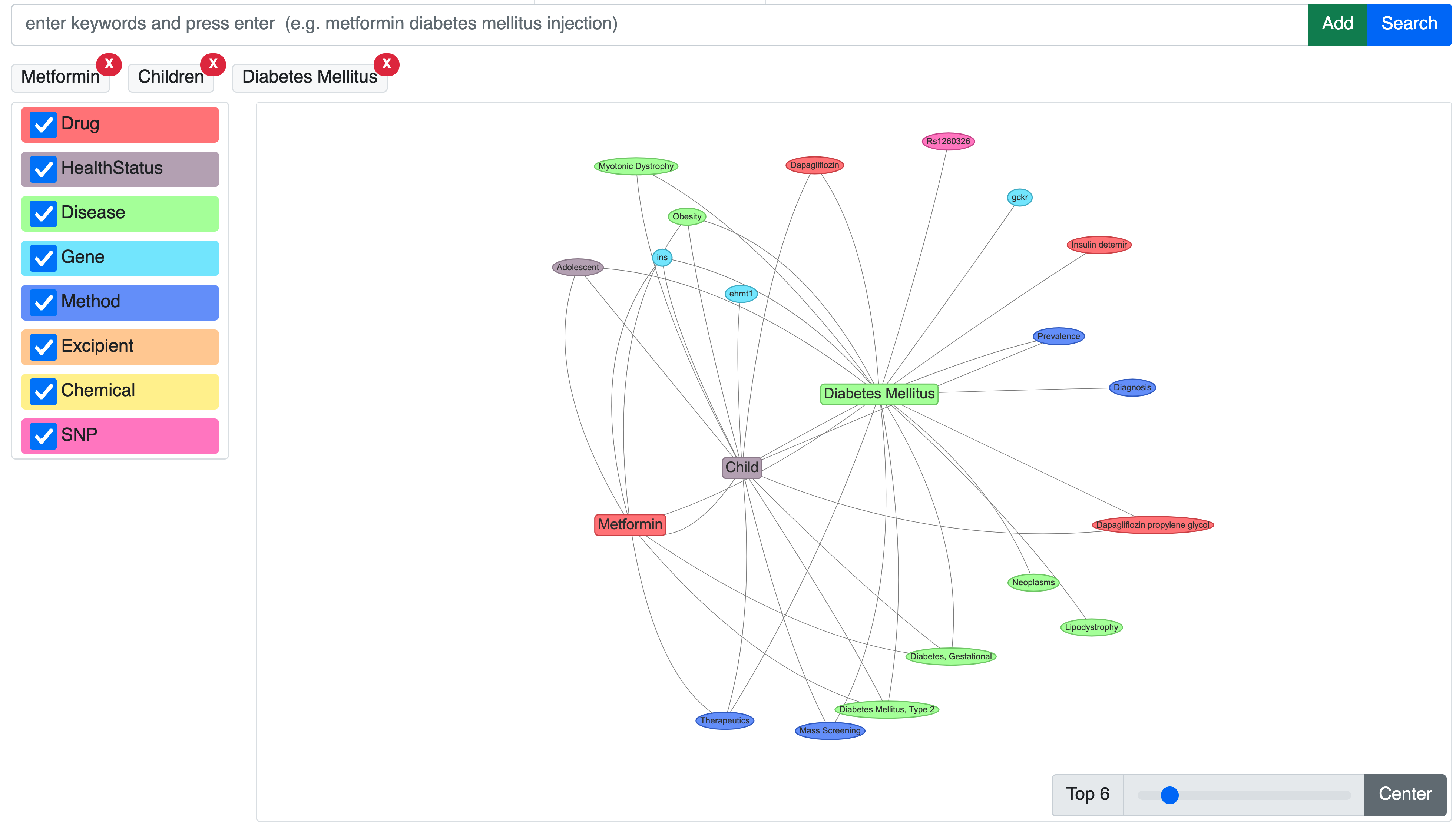}
    \caption{Pattern Visualization: The extracted pattern is shown in the center of the screen. Nodes are colored depending on their entity type. The searched entities are depicted as rectangle nodes with a larger font. Users may hide certain entity types or select how many edges are visualized at once.}
    \label{fig:pattern}
\end{figure}

\begin{figure}[t]
    \centering
    \includegraphics[trim={0 6cm 0 0}, clip, width=0.8\linewidth]{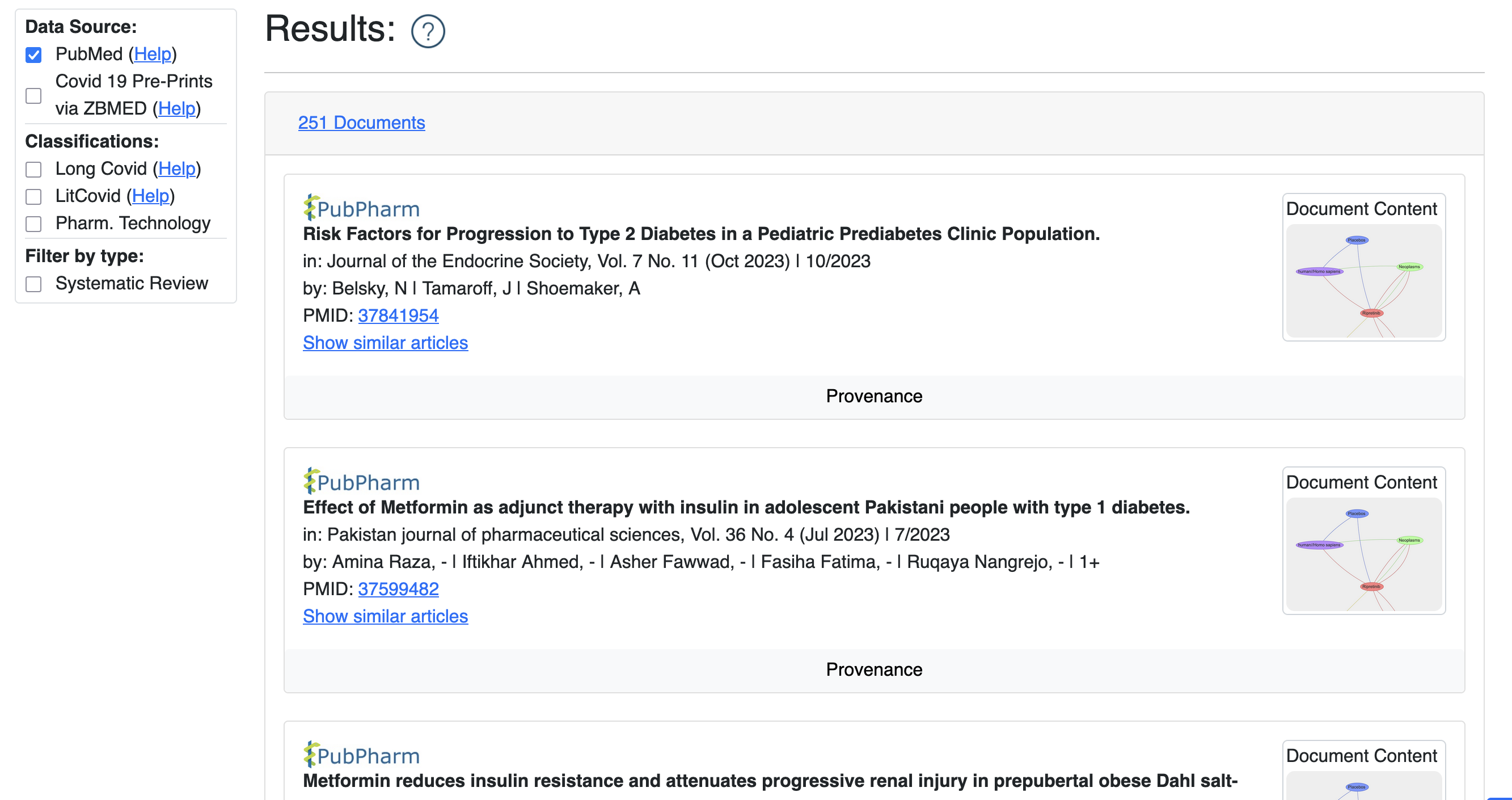}
    \caption{Result Lists: The system shows the document lists used to generate the narrative pattern. Filter options are available to further narrow down searches. }
    \label{fig:results}
\end{figure}

\section{Demonstration}
Our new narrative pattern discovery component has been integrated into our main discovery system\footnote{\url{https://beta.narrative.pubpharm.de}, tab \textit{Pattern Discovery (Beta)}}, and a tutorial video is available at\footnote{\url{https://pharmrxiv.de/receive/pharmrxiv_mods_00026752}}.
In the following, we first describe the user interface and explain our design decisions.
We then describe a preliminary user evaluation that we intend to extend in the future.

\subsection{User Interface}
Users can enter a list of searched entities by typing into a search bar. 
They are assisted with an autocompletion functionality that proposes known entity terms. 
They can then add entity terms to their search by pressing \texttt{enter} or clicking the \texttt{add} button on the right. 
Entities are then added to a list below the search bar. 
They can be removed by clicking on a red cancel sign. 
Next, users may start the search by pressing \texttt{enter} or clicking the \texttt{search} button.

The system replies with a color-coded graph representation. 
The searched entities are highlighted in the center of the visualization as rectangle nodes with a larger font. 
Every other node is a rounded oval.
Nodes' colors depend on their entity types, e.g., red for drugs.
A unified entity-type coloring is used across the whole discovery system.
We decided to keep the representation simple, so we removed edge labels. 
Suppose users want more details on a certain interaction (edge) between two entities. 
In that case, they can click on an edge and are forwarded to a corresponding search in our discovery system, i.e., literature is shown that supports the clicked interaction between two entities.
Entity types can be hidden or made visible via clicking the colored boxes on the left side of the screen. 
This way, users can narrow down the pattern to show only certain entity types, such as drugs, diseases, and targets.
Our document graph representation has already established a similar feature; see \autoref{fig:document}.
At the bottom, users may select how many edges should be visualized simultaneously. 
By default, this is set to five edges per concept.
If users scroll further down, they see a list of documents that contain the searched entities. 
This feature delivers Provenance, which allows users to screen the literature used to generate the pattern.
In addition, a source selection filter is shown on the left side.
Here, users can select which data sources are used for the pattern mining step.
They can narrow down their searches to specific collections or certain document classes, e.g., articles relevant to Pharmaceutical Technology or published within a specified time span. 
Screenshots of our user interface are shown in \autoref{fig:pattern} and \autoref{fig:results}.

\subsection{User Evaluation}
%\textbf{1 user evaluation}

\paragraph{Setup.}

For our small-scale user evaluation, we conducted one on one interviews taking around thirty minutes with five participants in German. Participants were familiar with PubPharm and the Narrative Service as they already took part in earlier studies.
Participants were able to access the tool via Zoom and remote desktop control. 
Two experienced interviewers lead the participants through the three-part study:
\textbf{1. Introduction.} (5 minutes) Participants were informed on the purpose and terms of the study before they consented to take part in the study. Afterwards, the interviewer gave a brief overview of the tool. One of the interviewers took notes while the other interviewer lead the participant through the study.
\textbf{2. Usage of the system with thinking aloud.} (15 minutes) Participants used the whole system freely while thinking aloud~\cite{lewisusing} to work on one of their research questions.
\textbf{3. Semi-structured interview.} (10 minutes) We followed an earlier evaluation on parts of the system~\cite{DBLP:conf/jcdl/KrollKSB23} for the semi-structured interview and asked participants the same questions on general thoughts regarding the tool, encountered problems, liked components, changes required for them to consider using the tool and if they had any other comments.

\paragraph{Results.}

% search
%
% entering keywords for search - s1 (not alone), s2, t, p
% BUT "you need to first understand what it wants then you are able to do it" - s2
%
% search vs add - s1
% search not clicked but seached - c
% search used keyword that was not entered - c, t - not mentioned
Participants \textbf{encountered problems} related to the \textit{search bar}: they did not immediately understand how keywords were supposed to be entered with one participant noting \textit{you need to first understand what it wants then you are able to do it}. 
They had trouble with selecting keywords the system had in the vocabulary.
The two buttons \texttt{search} and \texttt{add} were not descriptive enough. 
%
% clicked edges
%
% losing context - s1, p
% few results - t - not mentioned
% new tab and back to original query unclear - t
%
When \textit{clicking edges} in the displayed graph, participants were surprised that the triggered search would lose the context of the graph and only look for the entities associated with the edge (and not all entities that were searched in the first place). 
The click opened the results in a new tab leading to some difficulty to navigate back to the original graph.
%
% interface problems
%
% speaking error messages - s1, p
% size of symbols - s2, p
% similar colours - c
% not clear that fields are not click-able - t
% top x but nothing happens - c - bug - not mentioned
% problems scolling/zooming/clicking - c - due to zoom - not mentioned
%
Participants identified some \textit{interface problems} which are easy to correct from a technical point of view: cryptic error messages, small-sized symbols, similar colors used in the graph and uncertainty which fields can be clicked.
%
% additional observations
%
% unclear additional value - s1, t
% more time needed - t
% unclear scrolling down under graph or not used - s2, c
% no interaction with graph - s1
% no top x - s1
% no restriction of concept types - s1
We made some \textit{additional observations}: in general it seemed to have been unclear what the additional values was, that the new tool brought. Participants would have required more time. They often did not scroll to see the documents resulting the search but only checked out the graph, sometimes they did not interact with the graph, they did not use the \texttt{top x} functionality or the possibility to restrict the depicted concept types.

% few but fitting results for precise queries - s1, p
% good overview for diving into topic, intuitive - s2, c, p
% intuitive but one needs to play arount a bit - t, p
% logical if hovered over - s2
% able to evaluate results while searching, different from LLM - s2
As for \textbf{liked components} participants mentioned the fitting results for precise queries, the tool's power to provide a good overview of a topic, its intuitive and logical handling as well as the possibility to evaluate the results while searching which makes the tool stand out against using LLMs.
%
% interface 
%
% visualisation - s1, s2, c
% depiction of relations - s2, c
% new tab - s2, c, p
% interactions - c
The \textit{interface} was praised: its visualization of relations and keywords, the possible interactions as well as opening clicked edges in a new tab.
% 
% according to expectations
%
% clicking and showing literature - s1
% relation being focused when clicked - s2, t, p
% rotate, zoom - s2, c, t, p
%
For participants, the tool and especially the interaction with the graph was behaving \textit{according to their expectations}, clicking on an edge focuses on this specific part of the graph and shows literature backing the edge.
%
% observations
%
% clicking edges - s1, s2, c, t, p
% deleting keywords - s1, c, p
% able to restrict query results by using filter - c
% using top x - c, p
% clicking concepts on left hand side - c
% back to tab (recovery) - c, t, p
%
We \textit{observed} users clicking edges, deleting keywords and modifying or restricting their queries, using the \texttt{top x} functionality, restrict the depicted concept types in the graph, and recovering from involuntary clicks.  

% use without changes - s1 - not mentioned

% english terms - s1, s2, t
Participants mentioned some \textbf{required changes} or potential future features: 
Some participants were unsure how to express some technical terms in English which hints towards a language-independent query option.
% search
%
% excluding edges - s1
% unknown terms - s1, s2, c, t
% variables - c
% clear button - c
% sorting suggested keywords better - c
% typo resistant - s2
% search between 3 things wanted - c
% google query & wanting to pick pattern - s1
%
In terms of \textit{search} there are many ideas: exclusion of edges, inclusion of terms which are not part of the tool's vocabulary, the option to query with variables, a \texttt{clear all} button for queries, an improved sorting for suggested keywords, typo resistance, and enabling the search between three interconnected components.
%
% export
%
% export search/network - s2, c
% clipboard for interesting works - s2
Participants wished to \textit{export or save} their searches, networks and interesting works.
%
% visualisation/interface
%
% show pattern seaech and result list in one view - c
% feedback it's not my fault but the system's when clicking a field - t - not mentioned
% increase size of top x - not mentioned
% cross references - t
%
In terms of \textit{visualization} they wished for seeing the pattern search and result list in one view. One participant asked for the inclusion of cross-references.
%
% other
%
% video tutorial - c
% explain additional value - t
% say what clicking edge means - c, t
% would wish for combining data from pubmed + news - s2
%
\textit{Other} wishes were video tutorials, better explanations and an extension of the data source to also incorporate news articles.

\section{Conclusion}
In brief, this paper introduces a novel access path for PubPharm's Narrative Discovery System.
Our narrative pattern mining approach assists users with entity-driven exploratory search. 
This way users can explore the neighborhood of searched entities.
The conducted interviews verified the usefulness of our system and the entity-driven and network-based visualization. 

Future work will tackle described problems and improve the search. We intend to provide users with the option to enter keywords similar to a Google search before employing, e.g., an LLM to suggested potential interesting patterns.

\section*{Acknowledgments}
Supported by the Deutsche Forschungsgemeinschaft (DFG, German Research Foundation): PubPharm – the Specialized Information Service for Pharmacy (Gepris 267140244).

\bibliographystyle{splncs04}
\bibliography{references}

\end{document}